\documentclass[12pt,a4paper]{article}
\usepackage[a4paper,margin=1in,footskip=0.25in]{geometry}
\usepackage{multirow}
\usepackage{url}
\usepackage{graphicx}
\usepackage{subcaption}
\usepackage{amsmath}
\usepackage{undertilde}
\usepackage{colortbl,booktabs}
\usepackage{tikz}
\usetikzlibrary{shapes.geometric, arrows}
\tikzstyle{decision} = [diamond, draw, fill=white!15,text width=10em, text badly centered, node distance=1cm, inner sep=0pt]
\tikzstyle{block} = [rectangle, draw, fill=white!20, text width=6em, text centered, rounded corners, minimum height=1cm]
\tikzstyle{line} = [draw, -latex']
\tikzstyle{cloud} = [draw, ellipse,fill=white!10, node distance=3cm, minimum height=4em]

\begin{document}
	
	\begin{center}
		{\LARGE A modified risk detection approach of biomarkers by frailty effect on multiple time to event data }
	\end{center}
	\vspace{1em}
	
	\begin{center}
		{\textit{Atanu Bhattacharjee $^a$, Gajendra K. Vishwakarma $^b$, Souvik Banerjee $^{b}\footnote[1]{Corresponding Author}$}}	\vspace{1em}
		
		$^a$ Section of Biostatistics, Center for Cancer Epidemiology,
		
		Tata Memorial Center, Navi Mumbai-410210, India, 
		
		$^a$ Homi Bhabha National Institute, Mumbai, India
		
		e-mail : atanustat@gmail.com
		
		$^b$ Department of Mathematics \& Computing, Indian Institute of Technology Dhanbad, Dhanbad-826004, India, 
		
		e-mail: vishwagk@rediffmail.com, souvik.stat@gmail.com 
		
	\end{center}
	\vspace{2em}
	
	\begin{abstract}
		Multiple indications of disease progression found in a cancer patient by loco-regional relapse, distant metastasis and death. Early identification of these indications is necessary to change the treatment strategy. Biomarkers play an essential role in this aspect. The biomarkers can influence how particular cancer behaves and how it may respond to a specific treatment. The survival chance of a patient is dependent on the biomarker, and the treatment strategy also differs accordingly, e.g., the survival prediction of breast cancer patients diagnosed with HER2 positive status is different from the same with HER2 negative status. This results in a different treatment strategy.  So, the heterogeneity of the biomarker statuses or levels should be taken into consideration while modelling the survival outcome. This heterogeneity factor which is often unobserved, is called frailty. When multiple indications are present simultaneously, the scenario becomes more complex as only one of them can occur, which will censor the occurrence of other events. Incorporating independent frailties of each biomarker status for every cause of indications will not depict the complete picture of heterogeneity. The events indicating cancer progression are likely to be inter-related. So, the correlation should be incorporated through the frailties of different events. In our study, we considered a multiple events or risks model with a heterogeneity component. Based on the estimated variance of the frailty, the threshold levels of a biomarker are utilised as early detection tool of the disease progression or death. Additive-gamma frailty model is considered to account the correlation between different frailty components and estimation of parameters are performed using Expectation-Maximization Algorithm. This work is about handling multiple indications by frailty model and promote personalised medicine. With the extensive algorithm in R, we have obtained the threshold levels of activity of a biomarker in a multiple events scenario.
	\end{abstract}	
	
	
	\section{Introduction}\label{intro}   
	
	Identification of disease progression is an essential aspect in the treatment of cancer. It is observed that specific biological characteristics of the human body of a cancer patient deviate from the same of healthy people. Thus an accurate indicator which can capture this deviation at an early stage is required. This necessitates the search for measurable biological identities which can serve as the indicator of cancer onset. This type of biological identities is called as biomarkers \cite{califf2018}. The identification of this occurrences before-hand is necessary for treatment intervention \cite{Strimbu2010}. In survival analysis, the interest lies in the waiting time until the occurrence of a well-defined event. Based on the objective of studies, the event may be defined differently, such as distant metastasis(DM), loco-regional relapse(LR) and death. In some situations, more than one event may be observed, and the evolution of the patient's state over time is studied. Multi-state models play an essential role to analyse such complex biological processes\cite{bhattacharjee2020,putter2007}. Due to the inter-related nature of the states, individual analysis for each event will not reveal the dependency structure among them, and the contributions of the prognostic factors towards disease progression are misunderstood \cite{crowder2001}.
	
	In cancer treatment, biomarkers can influence how particular cancer behaves and how it may respond to a specific treatment. The survival probability of a patient depends on the status of the biomarker, and the treatment plan is also changed accordingly, e.g., the survival prediction of breast cancer patients diagnosed with HER2 positive status is different from the same with HER2 negative status. This results in a completely different treatment strategy as the treatment for HER2 positive has little chance to act as effectively as for HER2 negative. These different statuses or levels create a disparate effect on each of the events. Suppose, a biomarker is measured with continuous variables values $u_1,u_2,..., u_n$. The range of $u_i$'s can be split into two parts as $[\mbox{min} ~u_i,a]$ and $(a, \mbox{max}~ u_i]$ and thus the expression values can be divided into two levels. These two levels may have a different effect on disease progression as well as a treatment regime. The key lies in finding the optimal threshold level to visualise this heterogeneous effect more prominently. This heterogeneity between the cluster of observations which is often unobserved can be modelled using a random frailty term. The variance of the frailty term can be interpreted as a measure of heterogeneity between different groups of patients. Possible correlation between patients diagnosed with a specific biomarker level can be modelled using a shared frailty component. However, when multiple causes of event occurrence or competing risks are present simultaneously, using only one frailty term per biomarker level for every competing risk will not properly explain the overall heterogeneity. Similarly, individual frailties for each indication under each biomarker status fail to capture the dependency among the competing events. The frailty components among different causes of failures under a specific biomarker level are likely to be correlated. Thus, the overall heterogeneity can be modelled using a combined contribution of both level-specific and event-specific variabilities \cite{fiocco2009}. The construction of a prognostic tool for patient's treatment strategy by incorporating both of these variabilities is challenging in oncology research. 
	
	Due to the varying effect of biomarker statuses over the disease progression, proper detection of biomarker statuses carries enormous importance. The treatment procedures for different biomarker statuses are likely to be different. The gene expression collected for a particular biomarker is continuous in nature. It is usually assumed that all genes follow the same underlying distribution. However, researchers have found that the gene expression analysis of several cancer biomarkers show multimodal expression patters \cite{kippner2014, detorrente2020}. Genetic mutations and amplifications are often inconsistent across tumors and their effects can be visible in smaller subsets of patients. This provides the idea of bimodally and multi-modally expressed genes that have distinct modes corresponding to statistically significant difference in survival times\cite{moody2019}. Thus, it is required to obtain a proper way for classification of statuses based on some statistical measures. The frailty component can be used to serve this purpose. The patients with the same biomarker status are likely to be clustered, and hence frailty variance might be lower for them. Thus, with the combination of different threshold values, we can classify the biomarker statuses accordingly and investigate the frailty variance for each of these combinations to obtain the most useful classification of statuses. 
	In the context of medical studies, this event of interest defined in different forms as disease recurrence, progression etc. The widely used Cox proportional hazard model\cite{cox1972} describes the relationship between a set of explanatory variables and a time-to-event outcome. This model also takes into consideration of possible censoring mechanism when an event has not occurred during the study period. The observed event times are assumed to be independent for the application of this model. However, this assumption is not appropriate when the survival duration of patients with a common characteristic may have a possible correlation \cite{balan2020}. This type of dependence in survival data can be modelled using a random frailty term. It represents the unobserved heterogeneity effect between observations or between clusters of observations. The variance of frailty component can be interpreted as a measure of heterogeneity between the groups or individuals.
	
	Analogous to the single end-point analysis, the Cox proportional hazard model can be extended for competing events of interest with cause-specific hazards for each cause of failure. The focus of multi-state models is to study the process of moving from one state to another. The study of multiple risks with clustered data complicates the possible dependency structure. In the context of competing events, only one frailty term for all the causes of failures is not suitable to explain the overall randomness. Also, independent frailty terms for each cause of failures will ignore the dependency structure due to a particular risk. The random effect in the survival model was introduced by Beard \cite{beard1959} as a longevity factor for modeling human mortality. Vaupel et al. \cite{vaupel1979} first introduced the term frailty to indicate the unobservable random effect shared by subjects in a standard risk group. Keyfitz \& Littman \cite{keyfiz1979} developed a procedure to estimate the life expectancy taking into consideration the heterogeneity present in the population. Yashin et al.\cite{yashin1995} introduced the shared frailty model using gamma frailty to analyse twin survival data. Frailty models for bivariate survival data were proposed by Oakes\cite{oakes1989}. He also introduced several possible frailty models for the analysis of survival data. Extensive work on frailty models has been done for survival data in different areas of research\cite{aalen1988,hougaard1986,clayton1978,petersen1996}. 
	
	Fine and Gray's subdistribution hazard model\cite{fine1999} is useful for analysis of competing events. This model is also extended to incorporate a frailty component to account cluster dependencies on the cumulative incidence function of the event of interest in the presence of competing risks. The correlated frailty models in the presence of competing risks analysed by Wienke et al.\cite{wienke2002} with the assumption of independence between the risks. Liquet et al. \cite{liquet2012} investigated multi-state frailty models for multi-centre data using both independent and joint frailty for modelling the dependency structure among transition intensities. Rutten-Budde et al. \cite{rueten2019} analysed multi-centre heterogeneity with a competing risk frailty model for two diseases by considering both centre-specific and disease-specific random components. Abbring and van der Berg \cite{abbring2003} discussed on the identifiability of cause-specific hazard frailty models. Ha et al. presented a detailed h-likelihood-based inference procedure of a multivariate frailty model under competing risk setting \cite{doha2017}.
	
	In this paper, we considered a competing events model and incorporated biomarker level-dependent heterogeneity to observe its effect on the survival prediction. An additive gamma frailty model was used multiplicatively on the cause-specific hazard to model the dependency structure within biomarker levels and between two competing events. The expectation-maximisation algorithm utilised for estimation procedure, and a strategy outlined to compute the standard error of the estimates. Based on the frailty variance of the competing events at different biomarker levels, an ideal threshold level of the biomarker is defined, and risk prediction is performed in a more accurate manner. The obtained threshold levels of the biomarker will determine the most efficient risk prediction analogy for different competing events. In section 2, we discuss the proposed risk detection procedure. In section 3, the simulation study is described to show the efficacy of our procedure and effectivity in identifying the heterogeneity in risk. In section 4, we have applied the procedure on a real dataset, and corresponding results are presented.

	\section{Methodology}\label{methods}
	
	Suppose there are $n$ independent subjects (i.e., patients). Let, $T_i$ be the time to failure for $i^{th}$ patient, $i=1,2,...,n$ and $X_i$ is a $m \times 1$ vector of covariates. Let, $C_i$ denotes non-informative censoring time, independent of $T_i$. We assume the event indicator $\delta_i=1$ indicates the failure time is observed and $\delta_i=0$ if censored. The hazard function for the $i^{th}$ patient is defined as,
	\begin{equation}
	\lambda(t_i|X_{i})=\lim\limits_{\Delta t \rightarrow 0} Pr(T_i \in [t_i,t_i+\Delta t_i)|T_i\geq t_i,\textbf{X}_{i})/\Delta t_i, ~~~~~t_i>0
	\end{equation}
	The survival function $S(t)=Pr(T>t)$ denotes the probability of survival time being greater than $t$. For a patient died at $t_i$, its contribution to the likelihood function will be the density during that duration and it is written in terms of the products of survival function and hazard function i.e, $L_i=f(t_i)=S(t_i)\lambda(t_i)$. If the patient is still alive at $t_i$, then the likelihood will be formulated as $L_i=S(t_i)$. So, the survival likelihood for all the $n$ patients is defined as, 
	
	\begin{equation}
	{\displaystyle L=\prod_{i=1}^n [\lambda(t_i)]^{\delta_i}S(t_i)=\prod_{i=1}^n [\lambda(t_i)]^{\delta_i} ~exp~(-\Lambda(t))}
	\end{equation}    
	
	
	where $\Lambda(t)=\int_{0}^{t} \lambda(s)ds$. In the multi-state set-up, there is a correlation present between the diseases and the evolution of the patient's state from one to another is studied. Very often, the Markov assumption is adopted in the multi-state set-up for mathematical convenience\cite{liquet2012}. In our study, we ignore the transitions and only consider whichever event appears earlier. So, we obtain a competing risk framework with multiple risks that are present for a patient at the beginning of the study.
	
	In multiple risk analysis, the interest is to identify the death rate for a particular cause.  The hazard rate of a specific cause $j$ for patient $i$ is defined as,
	
	\begin{equation}
	\lambda_j(t_i)=\lim\limits_{\Delta t_i \to 0} \frac{Pr(t_i \leq T_i \leq t_i+\Delta t_i, \delta_i=j|T_i \geq t_i)}{\Delta t_i} 
	\end{equation}
	Here $\delta_i=1,2,...,J$ is the indicator of causes of failure, $\delta_i=0$ implies censored data, $T_i=\mbox{min} (T_{i1},T_{i2},...,T_{iJ},C_i)$ 
	
	Cumulative cause-specific hazard and the probability of no failure from any of the causes are   defined as,
	\begin{equation}
	\Lambda_j(t)=\int_{0}^{t} \lambda_j(s)ds ~~~~~~~~~~ \text{and,} ~~~~~~~~~~S(t)=exp~ \Big(-\sum_{j=1}^{J} \Lambda_j(t)\Big)
	\end{equation}
	
	
	So, under competing risks, the overall likelihood for all the patients with different causes of failures is defined as\cite{klein2016},
	\begin{equation}
	L=\prod_{i=1}^{n} exp~ \Big({-\sum_{j=1}^J \Lambda_j(t_i)}\Big) \prod_{j=1}^{J} \lambda_j(t_i)^{1_{\{\delta_i=j\}}}
	\end{equation}
	
	When the proportionality assumption is maintained, the widely used Cox proportional hazard (PH) model is useful to observe the effect of covariates on different causes of failures. Using the proportional hazard model, the cause-specific hazard function of cause $j$ for the $i^{th}$ patient is defined as, 
	\begin{equation}
	\lambda_j(t|X_i)=\lambda_{j0}(t) exp(\beta_j^T X_i)
	\end{equation}
	where $\lambda_{j0}$ is the cause-specific baseline hazard function, $\textbf{X}_i$ is the covariates for $i^{th}$ patient and $\pmb{\beta}_j$ is the corresponding coefficient for cause $j$. 
	
	Thus, the likelihood function for PH model using cause-specific hazards is defined as,
	\begin{equation}
	L(\pmb{\beta}|\pmb{X}_i)=\prod_{i=1}^{n}~exp~\Big(-\sum_j \Lambda_{j0}(t_i) exp(\pmb{\beta}_j^T \pmb{X}_i)\Big) \prod_{j=1}^{J} \Big[\lambda_{j0}(t_i)~exp~(\pmb{\beta}_j^T\pmb{X}_i)\Big]^{1_{\{\delta_i=j\} }}  
	\end{equation}
	
	
	The presence of frailty effect can not be ignored in survival data analysis. The effect of the unobserved heterogeneity over hazard function can be measured using a random component in survival models. The variance of the frailty term used as a measure to examine the heterogeneity in the data. The univariate frailty model introduced by Vaupel et al.\cite{vaupel1979} discussed the proportionality assumption in frailty components and thus introduced the idea of the multiplicative effect of frailty component over the hazard function. In individual frailty model, it assumed that each of the individuals has its frailty part; thus, there is the individual-level different impact on the hazard rate. In the shared frailty model, the frailty component shared among all the individuals belonging to a particular cluster or group. Different distributional assumptions are made for the frailty component namely Gamma distribution, Gaussian distribution, Exponential distribution, t-distribution etc. In this study, we assume that the expression of a biomarker can be split in different levels based on threshold values and those levels have heterogeneous effects on the survival outcome. The proportional hazard frailty model defined as,
	
	\begin{equation}
	\lambda(t|\pmb{X}_i,W_k)=W_k\lambda_0(t)~ exp~(\pmb{\beta}^T \pmb{X}_i)
	\end{equation}
	where $W_k$ is the level-specific or shared frailty effect of $k^{th}$ level of biomarker. In shared frailty, the frailty component is shared by all the individuals in a cluster. So, the frailty variance can be considered as a measure of heterogeneity among the patients in that cluster.
	
	
	Let, $\textbf{W}=(W_1,W_2,...,W_K)$ is the frailty effect associated with K levels of the biomarker. $d_k$ and $n_k$  are the number of events and patients  having specific $k^{th}$ level of a biomarker respectively. $\delta_{ki}=0, 1,2,...,J$ is the indicator of event/cause of failure for patient $i$ in level $k$. 
	\begin{equation}
	L(\pmb{\beta},\lambda_0|data,\textbf{W})=\prod_{k=1}^{K}f(W_k)\prod_{i=1}^{n_k}\Big( W_k \lambda_0(t_{ki})~exp(\pmb{\beta}^T\pmb{X}_{ki}) \Big)^{\delta_{ki}} ~exp~\Big(-W_k \Lambda_{0}(t_{ki})~exp~(\pmb{\beta}^T\pmb{X}_{ki})\Big)
	\end{equation}
	Here, the frailty component is assumed to follow Gamma distribution. The conjugate prior property of gamma distribution makes this a popular choice for frailty component given the survival data \cite{duchateau2007}. It benefits to compute the explicit form of the posterior distribution as well as the formulation of programming codes. Other distributional approaches of the frailty term can be found in the following literature \cite{balan2020}. 
	
	
	
	So, the likelihood of survival data with independent frailties for $J$ competing events is defined as,
	\begin{multline}
	L(\pmb{\beta}_1,\pmb{\beta}_2,...,\pmb{\beta}_J,\lambda_{10},\lambda_{20},...,\lambda_{J0}|data,\textbf{W})=\prod_{k=1}^{K} \Big(\prod_{j=1}^{J}f(W_{kj})\Big)\\ \prod_{i=1}^{n_k}~ exp~\Big(-\sum_j W_{kj} \Lambda_{j0}(t_{ki})~exp(\pmb{\beta}_j^T \pmb{X}_{ki})\Big)\\ \prod_{j=1}^{J} \Big(W_k \lambda_{j0}(t_{ki}) ~exp(\pmb{\beta}^T_j \pmb{X}_{ki}) \Big)^{\delta_{ki}}
	\end{multline}
	By integrating out the frailty terms, we obtain observed data likelihood \cite{hanagal2011}.
	
	
	Using same shared frailty irrespective of events is not enough to completely explain the randomness in survival data. The heterogeneity due to different events are also needed to be taken into account. So, if  we assign independent frailty terms corresponding to each cause of failures within a biomarker level, then the dependency due to effect of a particular level is ignored.  This correlation can be translated through the frailty components. Yashin et al.\cite{yashin1995} first proposed the decomposition of frailty term in two independent frailties, one of them being the shared frailty and the other is the cause specific independent frailty. Petersen et al. \cite{petersen1996} extended the Cox model to incorporate additive gamma frailties and estimated the parameters using the EM algorithm. 
	
	$W_{k1},W_{k2},...,W_{kJ} $ are frailty variables assigned to each cause of failure for a biomarker level $k$. So, for example, with 4 levels of a biomarker and 3 different causes of failures, the full set of frailty effects will be $\{W_{kj}\}_{\substack{k=1,...,4\\ j=1,...,3}}$. Each frailty component for level $k$ constructed adding two parts, one is a common component which has the same effect on all the causes, and another is the independent component which has unobserved cause-specific effects.  
	
	The frailties defined as, 
	
	\[W_{k1}=\frac{Z_{k0}+Z_{k1}}{\nu_0+\nu_1},~W_{k2}=\frac{Z_{k0}+Z_{k2}}{\nu_0+\nu_2},~\dots,~W_{kJ}=\frac{Z_{k0}+Z_{kJ}}{\nu_0+\nu_J}\]
	
	where, $Z_{k0}$ is the shared frailty component due to the levels of biomarkers and irrespective of the cause of failure. $Z_{k1}, Z_{k2},...,Z_{kJ}$ are risk-specific frailty components due to different biomarker levels. $Z_{k0}\sim \Gamma(\nu_0,1),~~Z_{k1}\sim \Gamma(\nu_1,1),~~\dots,~~ \dots,~~Z_{kJ}\sim \Gamma(\nu_J,1)$. $Z_{k0},Z_{k1},\dots, Z_{kJ}$ are independent frailty components for biomarker level $k$.
	Thus, the frailties will be distributed as, 
	\begin{equation}
	W_{k1}\sim \Gamma(\nu_0+\nu_1,\nu_0+\nu_1),~~W_{k2}\sim\Gamma(\nu_0+\nu_2,\nu_0+\nu_2),\dots,\dots,W_{kJ}\sim \Gamma(\nu_0+\nu_J,\nu_0+\nu_J)
	\end{equation}
	
	The variance and correlations will be defined as, 
	\begin{equation}
	Var(W_{k1})=\frac{1}{\nu_0+\nu_1}=\xi_1, Var(W_{k2})=\frac{1}{\nu_0+\nu_2}=\xi_2,~\dots,Var(W_{kJ})=\frac{1}{\nu_0+\nu_J}=\xi_J
	\end{equation}
	and
	\begin{equation}
	Corr(W_{kj_1},W_{kj_2})=\nu_0(\xi_{j_1} \xi_{j_2})^{1/2}=\rho_{j_1j_2}
	\end{equation}
	
	for two causes of failure $j_1$ and $j_2$ in a biomarker level $k$. The model parameters are estimated by maximising the logarithm of the likelihood function. The frailties within a biomarker level are associated with each other, and patients across different clusters are independent. So, the likelihood function is the product of the likelihoods of all patients across different levels. 
	
	The complete data likelihood for level $k$ given the data and frailty components will be 
	\begin{multline}
	L(\pmb{\beta}_1,\pmb{\beta}_2,...,\pmb{\beta}_J,\lambda_{10},\lambda_{20},...,\lambda_{J0}| data_k,Z_{k0},Z_{k1},Z_{k2},...,Z_{kJ})=\\ \prod_{i=1}^{n_k}\Big(\frac{Z_{k0}+Z_{k1}}{\nu_0+\nu_1}\lambda_{10}(t_{ki})exp(\pmb{\beta}_1^T\pmb{X}_{ki})\Big)^{1_{\{\delta_{ki}=1\}}}\\
	\Big(\frac{Z_{k0}+Z_{k2}}{\nu_0+\nu_2}\lambda_{20}(t_{ki})exp(\pmb{\beta}_2^T\pmb{X}_{ki})\Big)^{1_{\{\delta_{ki}=2\}}}
	\dots \dots \Big(\frac{Z_{k0}+Z_{kr}}{\nu_0+\nu_r}\lambda_{r0}(t_{ki})exp(\pmb{\beta}_r^T\pmb{X}_{ki})\Big)^{1_{\{\delta_{ki}=r\}}}\\ exp\Big(-\Big(\frac{Z_{k0}+Z_{k1}}{\nu_0+\nu_1}\Lambda_{10}(t_{ki})exp(\pmb{\beta}_1^T\pmb{X}_{ki})+\\ \frac{Z_{k0}+Z_{k2}}{\nu_0+\nu_2}\Lambda_{20}(t_{ki})exp(\pmb{\beta}_2^T\pmb{X}_{ki})+ \dots \dots+ \frac{Z_{k0}+Z_{kJ}}{\nu_0+\nu_J}\Lambda_{J0}(t_{ki})exp(\pmb{\beta}_J^T\pmb{X}_{ki})\Big)\Big)
	\label{likelihood}
	\end{multline}
	
	Here, $n_k$ denotes the number of patients in biomarker level $k$, $d_{kj}$ denotes the total number of patients with the cause of failure $j, j=1,2,...,J$. The baseline hazard function is remained unspecified which is estimated during the Expectation Maximization (EM) optimization. As the frailty components are not observed, so the estimation procedure becomes more complicated than the traditional maximum likelihood method. So, considering the unobserved frailties as missing information, we can estimate the model parameters using the EM algorithm. 	
	
	
	In genomic studies with a large number of gene expressions, it becomes necessary to combine the results of multiple tests and draw a necessary inference based on a meta-analysis. The p-value combination method is particularly advantageous when the data underlying the tests are difficult to merge. Results of multiple experiments are generally ordered based on a measure of association, such as p-value.  A single p-value may not reveal the proper association structure, especially when its value is close to the boundary level. Thus, statistical methods of associating multiple p-values to make a consolidated remark hold significant importance. 
	
	There are a number of p-value combination methods available in the literature with diverse statistical properties. Suppose, $p_1, p_2, ..., p_n$ are the p-values obtained from independent tests. If the underlying test statistics are denoted as $t_1, t_2, ..., t_n$ have absolutely continuous distribution under the corresponding null hypothesis, then the joint null hypothesis for the p-values is defined as,
	\begin{equation}
	H_0: p_i \sim Uniform[0,1],~~~i=1,2,...,n 
	\end{equation}  
	
	Several commonly used statistics are available for combining the p-values namely, $P_F=\sum_{i=1}^{n} log~p_i$ (Fisher\cite{fisher1934}), $P_P=-\sum_{i=1}^{n} log(1-p_i)$ (Pearson\cite{pearson1933}), $P_G=\sum_{i=1}^{n} log[p_i/(1-p_i)]$ (Mudholkar \& George\cite{mudholkar1979}), $P_E=\sum_{i=1}^{n} p_i$ (Edgington\cite{edgington1972}), $P_T=min(p_1,p_2,...,p_n)$ (Tippett\cite{tippett1931}). The combined p-value is $p^C=1-F(Y)$ where $Y$ is the statistic of the p-values. $Y=\sum_{i=1}^{n} T(p_i)$ where $T$ is the transformation performed on the p-values (This $T$ is different from the $T$ mentioned at the beginning of Section 2. Here $T$ is a function of p-values whereas earlier it was referred as the time to failure). Clearly, all of the combiners are monotonic in p-values and therefore optimal in some settings. 
	
	Prior to the modern computing era, the form of $T$ was considered based on the ease of computing power so that the joint distribution of the statistic can be evaluated easily. However, with the increase in computational strength, different forms of transformations are utilised using Monte Carlo simulation. The steps are as follows\cite{zaykin2007}: (1) Calculate $Y_0=\sum T(p_i)$ from the observed p-values, (2) Generate $Y_j=\sum T(p_i)$ for $j=1,2,...,M$ simulations, where $p_i \sim Uniform[0,1]$. The proportion of $Y_j \leq Y_0$ will give the combined p-values $p^C$. Extensions have been made to incorporate the dependency among p-values \cite{zaykin2002,dudbridge2004}.
	
	In our study to obtain ideal cutoffs for each of the biomarkers based on independent p-values, we employed different combiner methods to combine different p-values obtained from independent tests. For each gene, the gene expression values are divided into percentiles and taking each percentile value as threshold level, the gene expression is transformed into a binary variable. So, for a particular $r^{\text{th}}$ percentile $[r=1,2,...,99]$ of gene $g$, the expression $U_g$ is divided into two parts $[min~ U_{g},U_{gr}]$ and $[U_{gr},max~ U_{g}]$. $U_g$ denotes the expression of $g^{\text{th}}$ gene and $U_{gr}$ denotes the $r^{\text{th}}$ percentile of gene $g$. 
	
	Thus, we obtain \begin{equation}
	Q_{gr}=\begin{cases}
	0 ~~\textrm{for}~~ U_{g}<U_{gr}\\
	1 ~~\textrm{for}~~ U_{g}\geq U_{gr}\\ 
	\end{cases}
	\end{equation}
	
	For each of this cutoff values, the Cox proportional hazard model is fitted taking into account this binary variable along with other prognostic factors. For each of these equations, we obtain a corresponding p-value $p_{gk}$. Thus for genes $g=1,2,...,G$, the p-values are $p_{\utilde{1}}, p_{\utilde{2}},..., p_{\utilde{G}}$ where $p_{\utilde{g}}= \{p_{g1}, p_{g2},..., p_{g99}\}$ are the p-values for different threshold levels. 
	
	Here, the equation with the minimum of the p-values will provide the highest significance on the binary variable, i.e., it produces the highest significant level of the gene expression. This will depict the best cut-points for which the heterogeneity between the two groups is maximum. Thus, it will provide a more stringent threshold for better classification of gene expression.

	\section{Simulation}
	
	We performed a simulation study to generate and analyse datasets for assessing the performance of the correlated frailty model. The datasets have been made using different parameter values and obtain the effect of correlation due to varying combinations of the frailty components. Different parameter values considered for simulation, and based on the generated dataset, we performed the required analysis. 
	
\begin{figure}[!h]
		\centering
		\includegraphics[width=0.8\linewidth]{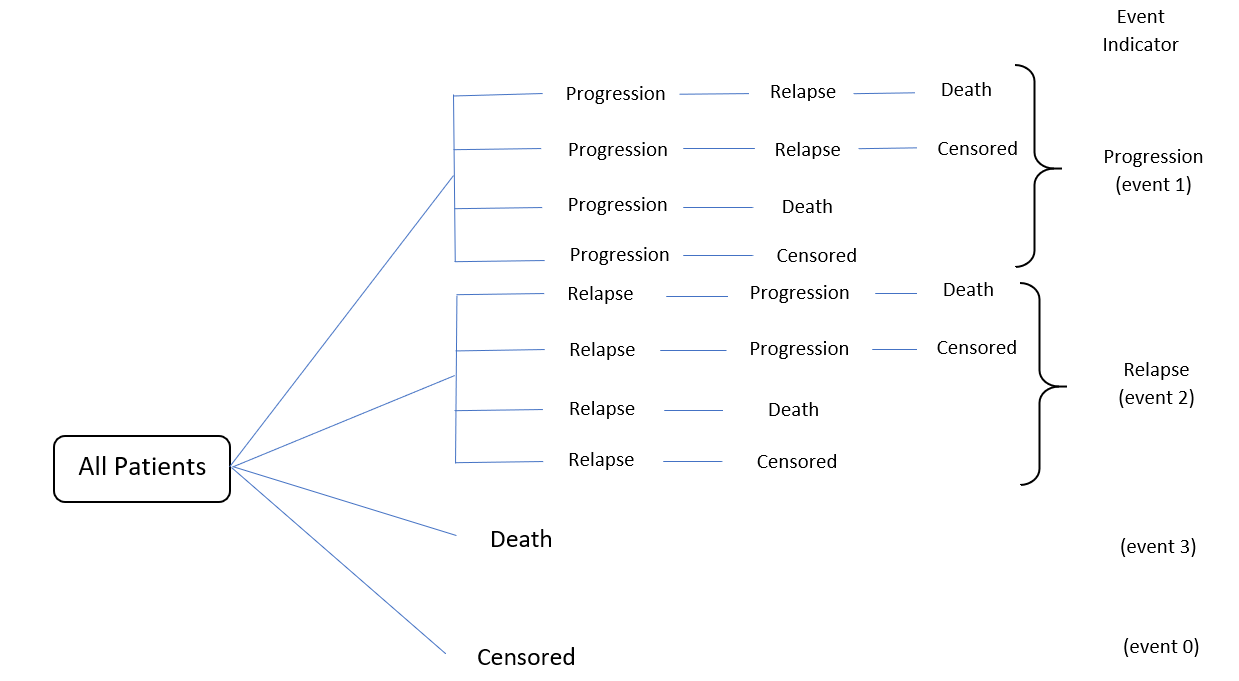} 
		\caption{Event indicator for competing risk model}
		\label{fig:flowchart}
	\end{figure}

	Our objective was to replicate the real-life scenario of cancer data obtained in the clinical study. We considered three different stages of HER2 for frailty component and three different events (Progression, Relapse and Death) for competing events. So, the value of $J$ is 3. For each event, we considered five different patients, so as a total, we obtained simulated data for 60 patients. The starting state regarded as the state of a patient while entering surgery and being alive with no evidence of disease post-surgery. The classification of events is described in figure \ref{fig:flowchart}.
	
	So, the frailty components are $W_{k1}, W_{k2}, W_{k3}$ in this analysis set-up. The shared frailty component is denoted as, $Z_{k0}$, which is biomarker level-specific frailty shared by all the risks. $Z_{k1}, Z_{k2}, Z_{k3}$ are the frailty components assigned for each of the risks.
	
	The frailty parameters are assumed to have a common scale as 1. The standard frailty components $Z_{k0}$ considered to have shape 1.5. The individual frailties $Z_{k1}, Z_{k2}, Z_{k3}$ have shape parameters 2, 2.5 and 3 respectively, to simulate the survival times and event statuses, the event times for three competing events are generated using Weibull distribution with scale parameters as 4.8, 5.2 and 5.5 respectively and shape parameters as 1.01, 1.02 and 1.04 respectively. Censoring times made using Exponential distributions with standard rate parameter 0.5. 
	
	Keeping in mind the real data scenario, we considered three different prognostic factors as covariates, namely, Age, t-stage and n-stage. The regression coefficients considered for the 3 covariates are $(-0.06, 0.5, 0.1)$, $(-0.05, 0.2, 0.2)$ and $(-0.03, 0.3, 0.3)$ respectively for 3 diseases. The Age considered as a random variable simulated using random number generation from Uniform distribution with lower bound ten and upper bound 70. The t-stage and n-stage are binary variables and randomly generated from 0 and 1. 0 considered for lower levels, and one is for higher levels. The event statuses for three competing events are denoted as 1, 2 and 3, whereas the event status for censored survival times denoted as 0.
	
	The data generation algorithm is explained in the following:
	
	\begin{itemize}
		\item[Step 1:] The frailty components $Z_{k0}, Z_{k1,...,Z_{k3}}$ are generated from gamma distributions.
		\item[Step 2:] Covariates are simulated from respective distributions.
		\item[Step 3:] Random numbers from uniform distribution are generated and the survival data is simulated for each of the events by inversing proportional hazard model with Weibull baseline hazard function.
		\item[Step 4:] Censoring times are generated from Exponential distribution.
		\item[Step 5:] The observed survival time is identified as the minimum of the survival times generated for 3 events and censoring time. The corresponding event status is also identified. 
	\end{itemize}
	
	\begin{table}[!t]
		\caption{Parameter estimates for Independent Frailty Model for Simulated Data}
		\begin{center}
			{\fontsize{8}{10} \selectfont
				\begin{tabular}{|c|c|c|c|c|c|c|c|c|c|}
					\hline
					parameter & Progression & & & Relapse  & & & Death & & \\
					\hline
					& true & estimate & std. & true & estimate & std.& true & estimate & std \\
					& value & estimate & error &  value & estimate & error & value & estimate & error\\
					\hline
					Age & -0.06 & -0.063 & 0.030 & -0.05 & -0.076 & 0.033 & -0.03 & -0.047 & 0.023\\
					t-stage &0.1 & 0.859 & 0.901 & 0.2  & -0.01 & 0.841 & 0.3 & 0.150 & 0.724 \\
					n-stage &0.5 & 0.192 & 0.916 & 0.2 & 0.632 & 0.837 & 0.3 & 0.282 & 0.723\\
					\hline
					Frailty & 0.25 & 7.7e-09 & p-value & 0.243 &  7.7e-09 & p-value &  0.238 & 7.5e-09 & p-value\\
					Variances & & &= 0.95 & &  &  = 0.99& & & = 0.96\\
					\hline
				\end{tabular}
			}
		\end{center}
		\label{table:simind1}
	\end{table}

	First, we fit an independent frailty model by taking into consideration the level effect of the biomarker separately on the competing event and ignoring the shared level-specific impact overall events. The results of the independent frailty model shown in table \ref{table:simind1}. It is evident the parameter estimates are heavily biased when the shared frailty ignored, and only cause-specific randomness considered. It fails to capture the correlation between the effects of the covariates. Also, the frailty variances are very low with high p-values which signifies no HER2 level effect on the risks. This independent frailty model is insufficient to analyse the simulated datasets. So, we need to opt for a model which will take into consideration both the individual and shared frailty effects.
	
	The results of the analysis are displayed in the following table \ref{table:simanals1}. The estimated correlations are also close to their actual values which validate the correlated frailty model. The results of the relationships shown in table \ref{table:simanals2}. The optimal regression estimates obtained using the Expectation Maximisation (EM) algorithm, which is tabulated exclusively for capturing the correlated frailty effect. The posterior distributions for each of the frailty parameters are classified using observed data likelihood.  
	
	From the table, we observe that the estimates of the fixed effect parameters for Age, n-stage and t-stage are close to the actual values and the estimates fall within the confidence intervals. The exact values of frailty variances also fall within the 95\% confidence intervals of the estimated values. The estimates are closer to the precise values for correlated frailty models than that of in the independent frailty models. 
	
	The standard errors for the coefficients of the covariates are generally smaller in correlated frailty model than the independent model. Also, the correlated frailty model can capture the shared frailty effects and the pairwise correlations among them. The estimated values are close to the valid parameter values, and the efficiency of the correlated frailty model explained based on the Empirical Standard Error (EmpSE) and Root Mean Squared Error (RMSE). The standard errors of the parameter estimates of the covariates in this model are lower than the independent frailty model. Also, the EmpSE and RMSE are much lower for the proposed correlated frailty model. The values of the RMSE and EmpSE of the frailty variances and correlations are also close to each other.
	
	\begin{table}[!h]
		\caption{Parameter estimates for Correlated frailty model on Simulated Data}
		\begin{center}
			{\fontsize{8}{10} \selectfont
				\begin{tabular}{|c|c|c|c|c|c|c|c|c|c|}
					\hline
					parameter & Progression & & & Relapse  & & & Death & & \\
					\hline
					& true & estimate & CI & true & estimate & CI & true & estimate & CI\\
					& value & estimate & CI &value & estimate & CI & value & estimate & CI\\
					\hline
					Age & -0.06 & -0.058 & (-0.114,-0.002)& -0.05 & -0.024 & (-0.063,0.015) & -0.03 & -0.023 & (-0.063,0.017)\\
					t-stage &0.1 & 0.290 & (-1.570,2.152) & 0.2  & 0.108 & (-1.172,1.387)& 0.3 & 0.067 & (-1.23,1.371) \\
					n-stage &0.5 & 0.191 & (-1.717,2.098) & 0.2 & 0.026 & (-1.265,1.318) & 0.3 & 0.102 & (-1.304,1.302)\\
					\hline
					Frailty & 0.286 & 0.346 & (0.315-0.371)& 0.25 &  0.299 & (0.266-0.310) &  0.222 & 0.268 & (0.220,0.296)\\
					Variances & & & EmpSE=0.271& &  & EmpSE=0.243 & & & EmpSE=0.239\\
					& & & RMSE=0.270& &  & RMSE=0.241 & & &RMSE=0.237\\
					\hline
				\end{tabular}
			}
		\end{center}
		\label{table:simanals1}
	\end{table}
	
	\begin{table}[!h]
		\caption{Frailty Correlations for Simulated Data}
		\begin{center}
			{\fontsize{8}{10} \selectfont
				\begin{tabular}{|c|c|c|c|c|c|c|c|c|c|c|c|}
					\hline
					Corr & true & estimate & CI & & true & estimate & CI & & true & estimate & CI\\
					-elations& value& & & & value & & & & value & &  \\
					\hline
					$\rho_{12}$ & 0.401 & 0.487 & (0.433,0.512) &$\rho_{13}$ & 0.377 & 0.490 & (0.462,0.522) &$\rho_{23}$ & 0.353 & 0.495 & (0.458,0.536)\\
					& & & EmpSE=0.307& & & & EmpMSE=0.334 & & & & EmpSE=0.330\\
					& & & RMSE=0.310& & & & RMSE=0.343 & & & &RMSE=0.351\\
					\hline
				\end{tabular}
			}
		\end{center}
		\label{table:simanals2}
	\end{table}
	
	\section{Application on breast cancer data}
	
	Analysis has been performed on a breast cancer dataset obtained from fresh frozen breast cancer tissue of every third patient diagnosed and treated between 1991 and 2004 at the Koo Foundation Sun-Yat-Sen Cancer Center. Patients whose the follow-up period is less than three years, other than the cause of death were excluded from the study. In case of non-availability of the designated sample, the following one was selected. A total of 447 samples were obtained out of which 135 samples were excluded due to insufficient RNA (n=1), poor RNA (n=116) or unacceptable micro-array quality (n=18). So, a total of 312 samples were eligible for the study. In addition to that, gene expression of additional 14 breast carcinoma samples was also selected, which were collected between 1999 and 2004. As a result, the dataset consists of disease information of 326 female patients. All the patients were treated according to the guidelines National Comprehensive Cancer Network. Following a modified mastectomy or breast-conserving surgery plus dissection of axillary nodes, the patients were given radiotherapy, adjuvant chemotherapy and/or, hormonal therapy. Neoadjuvant chemotherapy was provided to patients with locally advanced disease. This is an open-source data readily available in GEO database (GSE20685) and the study was performed with all necessary consents and ethics. Hence, we did not require any separate ethical approval or consent for our study.
	
	Breast cancer is one of the most common types of cancer diagnosed in females. The standard treatment protocol for breast cancer is surgery, followed by chemotherapy. A patient might experience progression of disease after surgery. In our study, they can either develop distant metastasis (Eventmet), i.e., tumour growth at a distant place or might experience death. The patients are observed for a period, and if they do not experience any of the event (Eventmet or death), their survival time is censored.
	
	\begin{figure}[!h]
		\centering
		\includegraphics[width=0.6\linewidth]{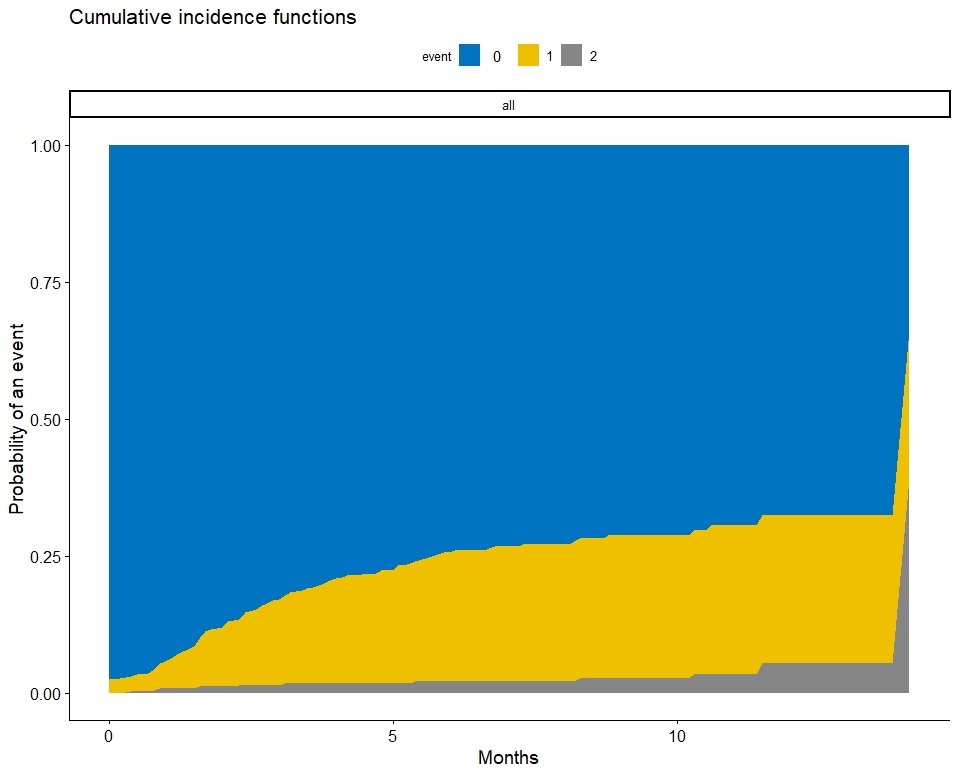} 
		\caption{Plot of cumulative incidence function for the events. event indicators: 0 = Censored, 1 = Distant Metastasis, 2 = Death}
		\label{fig:CIF}
	\end{figure}
	
	The competing risk model is explained in figure \ref{fig:flowchart}. The event of interest is distant metastasis and death is considered as a competing event. The starting date is the time when a patient enters the study after removing the tumour. To construct the competing risk model, we considered the event whichever occurs first out of distant metastasis and death. The choice of covariates considered for the study are age (continuous variable), t-stage(categorical variable; 4 categories: 1, 2, 3 \& 4), m-stage (categorical variable; 4 categories: 0, 1, 2 \& 3) and n-stage (categorical variable; 5 categories: 0, 1, 2, 3 \& 4). The data also consists of expression values for four biomarkers HER2 (Human Epidermal Growth Factor Receptor 2), ER (Estrogen Receptor), PR (Progesterone receptor) \& ERBB2 (Erb-b2 Receptor Tyrosine Kinase 2). The expression values of these genes are continuous variables and interest is to obtain optimal threshold values of the gene expressions to explore the effect of the partitions on survival prediction. The cumulative incidence functions for both DM and death are plotted in Figure \ref{fig:CIF}
	
	\begin{table}[!h]
		\caption{Competing Risk Model to show the effect of covariates}
		\begin{center}
			\begin{tabular}{|l|l|l|l|l|l|l|}
				\hline
				\multirow{2}{*}{variable} &      \multicolumn{3}{c|}{DM}  &    \multicolumn{3}{c|}{Death}          \\
				\cline{2-7}
				& HR      & std. error & p-value & HR       & std. error & p-value \\
				\hline
				age      & 0.998   & 1.012      & 0.856   & 1.025    & 1.030      & 0.401   \\
				tstage1  & 1       &            &         & 1        &            &         \\
				tstage2  & 0.552   & 1.433      & 0.099   & 0.173    & 2.372      & 0.042   \\
				tstage3  & 0.657   & 1.554      & 0.341   & 2.495    & 2.800      & 0.374   \\
				tstage4  & 1.667   & 1.348      & 0.088   & 1.342    & 3.169      & 0.799   \\
				mstage0  & 1       &            &         & 1        &            &         \\
				mstage1  & 14.77   & 1.482      & 0.000   & 6.92e-09 & Inf        & 0.998   \\
				mstage2  & 15.73   & 1.534      & 0.000   & 5.23e-09 & Inf        & 0.998   \\
				mstage3  & 27.12   & 1.528      & 0.000   & 8.41e-09 & Inf        & 0.998   \\
				nstage0  & 1       &            &         & 1        &            &         \\
				nstage1  & 1.06e08 & Inf        & 0.994   & 2.034    & 2.296      & 0.393   \\
				nstage2  & 1.63e08 & Inf        & 0.994   & 3.611    & 2.400      & 0.142   \\
				nstage3  & 1.57e08 & Inf        & 0.994   & 1.854    & 3.835      & 0.646   \\
				nstage4  & 3.39e08 & Inf        & 0.994   & 1.594    & Inf        & 1.000  \\
				\hline
			\end{tabular}
		\end{center}
		\label{table:datacomp}
	\end{table}

	Firstly, we performed a simple cause-specific regression analysis on survival data using Cox proportional hazard model. The effect of the corresponding covariates on the hazard function are shown in table \ref{table:datacomp}. From the table, we can see that for the t-stage level 2 and 3, the hazard ratio with respect to level 1 is less than 1 indicating a decrease in event occurrence, whereas for t-stage level 4, the hazard rate increases. For m-stage level 1,2,3, the hazard ratios are very high whereas for n-stage levels, the p-values are quite high. For death, the hazard ration for t-stage level 3 and 4 are more than 1 with respect to level 1. For the m-stage levels, the p-values are close to 1. However, for the sake of biological relevance, we have not discarded any of those covariates. We have taken one gene at a time and transformed the gene expression values into a binary variable. The percentiles of the gene expression are considered as different threshold values and analysis is performed for each of this partitions to obtain the frailty variance using correlated frailty model. Our interest to obtain the optimal threshold level for which the biomarker levels have the most distinct effect on the hazard ratio. The higher frailty variance will portray more significant heterogeneity between the two groups, thus more distinct effect on hazard ratios. 
	
	The threshold values which produces maximum frailty variance for HER2, ER, PR, ERBB2 are 10.713, 8.538, 8.511, 10.584, respectively. The frailty variances depict the heterogeneity in between two groups. Also, it is well known that p-value gives evidence against the null hypothesis. Lower the p-value, higher the significance of a variable in a model. So, if we transform the gene expressions into a binary variable by partitioning on different threshold values and use Cox proportional hazard model, then the equation with minimum p-value gives maximum importance on the binary variable that produces a highest significant level of gene expression. As a result, the p-values and frailty variances computed for different threshold values will be negatively correlated. The correlation between p-values and frailty variances for different threshold values are -0.33,-0.44,-0.56 and -0.32 for HER2, PR, ER and ERBB2 respectively. 
	
	\begin{table}[!h]
		\caption{Threshold values obtained for different genes. The genes are selected in the following order. The starting values for threshold are 1st(Q1), 2nd(Q2) \& 3rd(Q3) quartile respectively. Cutoff1, cutoff2, cutoff3 \& cutoff4 denotes the optimal cutoffs obtained for the genes.}
		\begin{center}
			{\fontsize{6}{12} \selectfont
				\begin{tabular}{|l|llll|llll|llll|}
					\hline
					\multirow{2}{*}{gene order} &      \multicolumn{4}{c|}{Starting threshold value = Q1}  &    \multicolumn{4}{c|}{Starting threshold value = Q2} & \multicolumn{4}{c|}{Starting threshold value = Q3}   \\
					\cline{2-13}
					& cutoff1 & cutoff2 & cutoff3 & cutoff4 & cutoff1 & cutoff2 & cutoff3 & cutoff4 & cutoff1 & cutoff2 & cutoff3 & cutoff4 \\
					\hline
					ER, ERBB2, HER2, PR    & 11.655  & 11.077  & 11.046  & 12.559  & 11.655  & 11.077  & 10.913  & 12.559  & 12.750  & 11.077  & 11.046  & 8.511   \\
					ER, ERBB2, PR, HER2  & 11.655  & 11.077  & 12.559  & 11.046  & 11.655  & 11.077  & 12.559  & 10.779  & 11.655  & 11.077  & 12.559  & 10.913  \\
					ER, HER2, ERBB2, PR    & 11.655  & 11.046  & 10.535  & 12.559  & 11.655  & 10.913  & 11.077  & 12.559  & 12.750  & 11.046  & 11.077  & 8.511   \\
					ER, HER2, PR, ERBB2 & 11.655  & 11.046  & 12.559  & 11.077  & 11.655  & 10.913  & 12.559  & 11.077  & 12.750  & 11.046  & 8.511   & 11.077  \\
					ER, PR, ERBB2, HER2  & 11.655  & 12.559  & 11.077  & 11.046  & 11.655  & 14.126  & 11.077  & 10.779  & 11.655  & 12.559  & 11.077  & 10.913  \\
					ER, PR, HER2, ERBB2 & 11.655  & 14.126  & 10.779  & 11.077  & 11.655  & 14.126  & 10.779  & 11.077  & 11.655  & 8.511   & 10.713  & 11.077  \\
					ERBB2, ER, HER2, PR    & 11.077  & 11.655  & 11.046  & 12.559  & 11.077  & 11.655  & 10.913  & 12.559  & 11.077  & 12.750  & 11.046  & 8.511   \\
					ERBB2, ER, PR, HER2  & 11.077  & 11.655  & 12.559  & 11.046  & 11.077  & 11.655  & 12.559  & 10.779  & 11.077  & 11.655  & 12.559  & 10.913  \\
					ERBB2, HER2, ER, PR    & 11.077  & 10.913  & 11.655  & 12.559  & 11.077  & 10.913  & 11.655  & 12.559  & 11.077  & 11.046  & 11.655  & 8.511   \\
					ERBB2, HER2, PR, ER    & 11.077  & 10.913  & 12.559  & 11.655  & 11.077  & 11.046  & 8.511   & 11.655  & 11.077  & 11.046  & 8.511   & 11.655  \\
					ERBB2, PR, ER, HER2  & 11.077  & 12.559  & 11.655  & 11.046  & 11.077  & 14.126  & 11.655  & 10.779  & 11.077  & 12.559  & 11.655  & 10.913  \\
					ERBB2, PR, HER2, ER    & 11.077  & 12.559  & 10.913  & 11.655  & 11.077  & 12.559  & 10.913  & 11.655  & 11.077  & 12.559  & 10.913  & 11.655  \\
					HER2, ER, ERBB2, PR    & 10.913  & 11.655  & 10.535  & 12.559  & 10.913  & 11.655  & 11.077  & 12.559  & 10.713  & 11.655  & 11.077  & 8.511   \\
					HER2, ER, PR, ERBB2 & 10.913  & 11.655  & 12.559  & 11.077  & 10.913  & 11.655  & 12.559  & 11.077  & 10.713  & 11.655  & 8.511   & 11.077  \\
					HER2, ERBB2, ER, PR    & 10.913  & 11.077  & 11.655  & 12.559  & 10.913  & 11.077  & 11.655  & 12.559  & 10.713  & 11.077  & 11.655  & 8.511   \\
					HER2, ERBB2, PR, ER    & 10.913  & 11.077  & 12.559  & 11.655  & 10.713  & 11.077  & 8.511   & 11.655  & 10.713  & 11.077  & 8.511   & 11.655  \\
					HER2, PR, ER, ERBB2 & 10.713  & 14.126  & 11.655  & 11.077  & 10.713  & 14.126  & 11.655  & 11.077  & 10.713  & 14.126  & 11.655  & 11.077  \\
					HER2, PR, ERBB2, ER    & 10.713  & 14.126  & 11.077  & 11.655  & 10.713  & 14.126  & 11.077  & 11.655  & 10.713  & 14.126  & 11.077  & 11.655  \\
					PR, ER, ERBB2, HER2  & 12.559  & 11.655  & 11.077  & 11.046  & 14.126  & 11.655  & 11.077  & 10.779  & 12.559  & 11.655  & 11.077  & 10.913  \\
					PR, ER, HER2, ERBB2 & 14.126  & 11.655  & 10.779  & 11.077  & 14.126  & 11.655  & 10.779  & 11.077  & 14.126  & 11.655  & 10.713  & 11.077  \\
					PR, ERBB2, ER, HER2  & 12.559  & 11.077  & 11.655  & 11.046  & 14.126  & 11.077  & 11.655  & 10.779  & 12.559  & 11.077  & 11.655  & 10.913  \\
					PR, ERBB2, HER2, ER    & 12.559  & 11.077  & 10.913  & 11.655  & 12.559  & 11.077  & 10.913  & 11.655  & 12.559  & 11.077  & 10.913  & 11.655  \\
					PR, HER2, ER, ERBB2 & 14.126  & 10.713  & 11.655  & 11.077  & 14.126  & 10.713  & 11.655  & 11.077  & 14.126  & 10.713  & 11.655  & 11.077  \\
					PR, HER2, ERBB2, ER    & 14.126  & 10.713  & 11.077  & 11.655  & 14.126  & 10.713  & 11.077  & 11.655  & 14.126  & 10.713  & 11.077  & 11.655 \\
					\hline
				\end{tabular}
			}
		\end{center}
		\label{table:genethreshold}
	\end{table}
	
	Different studies have shown significant association among the biomarkers related to breast cancer\cite{siadati2015,devi2015}. The dependency among biomarker influences the selection of optimal threshold value of gene expressions. So, it will be erroneous to independently identify the threshold values based on p-values for each of the genes using a binary variable. To overcome this issue, one can incorporate all the binary variables generated from every biomarker and perform the analysis based on the Cox proportional hazard model. However, this approach also suffers the problem of multiple testing, which increases the chance of false positives. For example, if the range of a particular gene expression is divided into 100 equal parts and each of them is considered as a threshold value, then for k biomarkers, there will be $100^k$ threshold values considering all possible combinations of threshold values of the biomarkers. Thus, performing this analysis becomes difficult without enough computation power. 
	
	\begin{table}[!h]
		\caption{Shared Gamma Frailty variances using proportional hazard model for patients having gene expressions above and below of the threshold values. lower\textunderscore fvar denotes frailty variance for patients with gene expression value lower than threshold level and upper\textunderscore fvar denotes frailty variance for patients with gene expression value higher than threshold level. The frailty variances are corresponding to the threshold values obatined in Table 5- with starting value as Q2.}
		\begin{center}
			{\fontsize{6}{12} \selectfont
				\begin{tabular}{|l|ll|l|ll|l|ll|l|ll|}
					\hline
					gene1 & lower\_fvar & upper\_fvar & gene2 & lower\_fvar & upper\_fvar & gene3 & lower\_fvar & upper\_fvar & gene4 & lower\_fvar & upper\_fvar \\
					\hline
					ER    & 5.00E-09    & 5.00E-09    & ERBB2 & 5.00E-09    & 5.11E-07    & HER2  & 0.183695    & 5.00E-09    & PR    & 0.264338    & 5.00E-11    \\
					ER    & 5.00E-09    & 5.00E-09    & ERBB2 & 5.00E-09    & 5.11E-07    & PR    & 0.264338    & 5.00E-11    & HER2  & 4.60E-05    & 5.00E-09    \\
					ER    & 5.00E-09    & 5.00E-09    & HER2  & 0.183695    & 5.00E-09    & ERBB2 & 5.00E-09    & 5.11E-07    & PR    & 0.264338    & 5.00E-11    \\
					ER    & 5.00E-09    & 5.00E-09    & HER2  & 0.183695    & 5.00E-09    & PR    & 0.264338    & 5.00E-11    & ERBB2 & 5.00E-09    & 5.11E-07    \\
					ER    & 5.00E-09    & 5.00E-09    & PR    & 5.00E-09    & 5.00E-13    & ERBB2 & 5.00E-09    & 5.11E-07    & HER2  & 4.60E-05    & 5.00E-09    \\
					ER    & 5.00E-09    & 5.00E-09    & PR    & 5.00E-09    & 5.00E-13    & HER2  & 4.60E-05    & 5.00E-09    & ERBB2 & 5.00E-09    & 5.11E-07    \\
					ERBB2 & 5.00E-09    & 5.11E-07    & ER    & 5.00E-09    & 5.00E-09    & HER2  & 0.183695    & 5.00E-09    & PR    & 0.264338    & 5.00E-11    \\
					ERBB2 & 5.00E-09    & 5.11E-07    & ER    & 5.00E-09    & 5.00E-09    & PR    & 0.264338    & 5.00E-11    & HER2  & 4.60E-05    & 5.00E-09    \\
					ERBB2 & 5.00E-09    & 5.11E-07    & HER2  & 0.183695    & 5.00E-09    & ER    & 5.00E-09    & 5.00E-09    & PR    & 0.264338    & 5.00E-11    \\
					ERBB2 & 5.00E-09    & 5.11E-07    & HER2  & 0.696243    & 5.00E-09    & PR    & 5.00E-09    & 5.00E-09    & ER    & 5.00E-09    & 5.00E-09    \\
					ERBB2 & 5.00E-09    & 5.11E-07    & PR    & 5.00E-09    & 5.00E-13    & ER    & 5.00E-09    & 5.00E-09    & HER2  & 4.60E-05    & 5.00E-09    \\
					ERBB2 & 5.00E-09    & 5.11E-07    & PR    & 0.264338    & 5.00E-11    & HER2  & 0.183695    & 5.00E-09    & ER    & 5.00E-09    & 5.00E-09    \\
					HER2  & 0.183695    & 5.00E-09    & ER    & 5.00E-09    & 5.00E-09    & ERBB2 & 5.00E-09    & 5.11E-07    & PR    & 0.264338    & 5.00E-11    \\
					HER2  & 0.183695    & 5.00E-09    & ER    & 5.00E-09    & 5.00E-09    & PR    & 0.264338    & 5.00E-11    & ERBB2 & 5.00E-09    & 5.11E-07    \\
					HER2  & 0.183695    & 5.00E-09    & ERBB2 & 5.00E-09    & 5.11E-07    & ER    & 5.00E-09    & 5.00E-09    & PR    & 0.264338    & 5.00E-11    \\
					HER2  & 5.00E-11    & 5.00E-09    & ERBB2 & 5.00E-09    & 5.11E-07    & PR    & 5.00E-09    & 5.00E-09    & ER    & 5.00E-09    & 5.00E-09    \\
					HER2  & 5.00E-11    & 5.00E-09    & PR    & 5.00E-09    & 5.00E-13    & ER    & 5.00E-09    & 5.00E-09    & ERBB2 & 5.00E-09    & 5.11E-07    \\
					HER2  & 5.00E-11    & 5.00E-09    & PR    & 5.00E-09    & 5.00E-13    & ERBB2 & 5.00E-09    & 5.11E-07    & ER    & 5.00E-09    & 5.00E-09    \\
					PR    & 5.00E-09    & 5.00E-13    & ER    & 5.00E-09    & 5.00E-09    & ERBB2 & 5.00E-09    & 5.11E-07    & HER2  & 4.60E-05    & 5.00E-09    \\
					PR    & 5.00E-09    & 5.00E-13    & ER    & 5.00E-09    & 5.00E-09    & HER2  & 4.60E-05    & 5.00E-09    & ERBB2 & 5.00E-09    & 5.11E-07    \\
					PR    & 5.00E-09    & 5.00E-13    & ERBB2 & 5.00E-09    & 5.11E-07    & ER    & 5.00E-09    & 5.00E-09    & HER2  & 4.60E-05    & 5.00E-09    \\
					PR    & 0.264338    & 5.00E-11    & ERBB2 & 5.00E-09    & 5.11E-07    & HER2  & 0.183695    & 5.00E-09    & ER    & 5.00E-09    & 5.00E-09    \\
					PR    & 5.00E-09    & 5.00E-13    & HER2  & 5.00E-11    & 5.00E-09    & ER    & 5.00E-09    & 5.00E-09    & ERBB2 & 5.00E-09    & 5.11E-07    \\
					PR    & 5.00E-09    & 5.00E-13    & HER2  & 5.00E-11    & 5.00E-09    & ERBB2 & 5.00E-09    & 5.11E-07    & ER    & 5.00E-09    & 5.00E-09 \\
					\hline
				\end{tabular}
			}
		\end{center}
		\label{table:gammafvar}
	\end{table}
	
	\begin{table}[!h]
		\caption{Shared Gaussian Frailty variances using proportional hazard model for patients having gene expressions above and below of the threshold values. lower\textunderscore fvar denotes frailty variance for patients with gene expression value lower than threshold level and upper\textunderscore fvar denotes frailty variance for patients with gene expression value higher than threshold level. The frailty variances are corresponding to the threshold values obatined in Table 5- with starting value as Q2.}
		\begin{center}
			{\fontsize{6}{12} \selectfont
				\begin{tabular}{|l|ll|l|ll|l|ll|l|ll|}
					\hline
					gene1 & lower\_fvar & upper\_fvar & gene2 & lower\_fvar & upper\_fvar & gene3 & lower\_fvar & upper\_fvar & gene4 & lower\_fvar & upper\_fvar \\
					\hline
					ER    & 6.29791     & 2.028077    & ERBB2 & 3.023438    & 3.159443    & HER2  & 0.520574    & 2.634373    & PR    & 2.248102    & 2.996094    \\
					ER    & 6.29791     & 2.028077    & ERBB2 & 3.023438    & 3.159443    & PR    & 2.248102    & 2.996094    & HER2  & 1.082031    & 2.267933    \\
					ER    & 6.29791     & 2.028077    & HER2  & 0.520574    & 2.634373    & ERBB2 & 3.023438    & 3.159443    & PR    & 2.248102    & 2.996094    \\
					ER    & 6.29791     & 2.028077    & HER2  & 0.520574    & 2.634373    & PR    & 2.248102    & 2.996094    & ERBB2 & 3.023438    & 3.159443    \\
					ER    & 6.29791     & 2.028077    & PR    & 2.116123    & 0.000651    & ERBB2 & 3.023438    & 3.159443    & HER2  & 1.082031    & 2.267933    \\
					ER    & 6.29791     & 2.028077    & PR    & 2.116123    & 0.000651    & HER2  & 1.082031    & 2.267933    & ERBB2 & 3.023438    & 3.159443    \\
					ERBB2 & 3.023438    & 3.159443    & ER    & 6.29791     & 2.028077    & HER2  & 0.520574    & 2.634373    & PR    & 2.248102    & 2.996094    \\
					ERBB2 & 3.023438    & 3.159443    & ER    & 6.29791     & 2.028077    & PR    & 2.248102    & 2.996094    & HER2  & 1.082031    & 2.267933    \\
					ERBB2 & 3.023438    & 3.159443    & HER2  & 0.520574    & 2.634373    & ER    & 6.29791     & 2.028077    & PR    & 2.248102    & 2.996094    \\
					ERBB2 & 3.023438    & 3.159443    & HER2  & 2.254859    & 3.161833    & PR    & 3.027314    & 2.357232    & ER    & 6.29791     & 2.028077    \\
					ERBB2 & 3.023438    & 3.159443    & PR    & 2.116123    & 0.000651    & ER    & 6.29791     & 2.028077    & HER2  & 1.082031    & 2.267933    \\
					ERBB2 & 3.023438    & 3.159443    & PR    & 2.248102    & 2.996094    & HER2  & 0.520574    & 2.634373    & ER    & 6.29791     & 2.028077    \\
					HER2  & 0.520574    & 2.634373    & ER    & 6.29791     & 2.028077    & ERBB2 & 3.023438    & 3.159443    & PR    & 2.248102    & 2.996094    \\
					HER2  & 0.520574    & 2.634373    & ER    & 6.29791     & 2.028077    & PR    & 2.248102    & 2.996094    & ERBB2 & 3.023438    & 3.159443    \\
					HER2  & 0.520574    & 2.634373    & ERBB2 & 3.023438    & 3.159443    & ER    & 6.29791     & 2.028077    & PR    & 2.248102    & 2.996094    \\
					HER2  & 1.753906    & 2.369918    & ERBB2 & 3.023438    & 3.159443    & PR    & 3.027314    & 2.357232    & ER    & 6.29791     & 2.028077    \\
					HER2  & 1.753906    & 2.369918    & PR    & 2.116123    & 0.000651    & ER    & 6.29791     & 2.028077    & ERBB2 & 3.023438    & 3.159443    \\
					HER2  & 1.753906    & 2.369918    & PR    & 2.116123    & 0.000651    & ERBB2 & 3.023438    & 3.159443    & ER    & 6.29791     & 2.028077    \\
					PR    & 2.116123    & 0.000651    & ER    & 6.29791     & 2.028077    & ERBB2 & 3.023438    & 3.159443    & HER2  & 1.082031    & 2.267933    \\
					PR    & 2.116123    & 0.000651    & ER    & 6.29791     & 2.028077    & HER2  & 1.082031    & 2.267933    & ERBB2 & 3.023438    & 3.159443    \\
					PR    & 2.116123    & 0.000651    & ERBB2 & 3.023438    & 3.159443    & ER    & 6.29791     & 2.028077    & HER2  & 1.082031    & 2.267933    \\
					PR    & 2.248102    & 2.996094    & ERBB2 & 3.023438    & 3.159443    & HER2  & 0.520574    & 2.634373    & ER    & 6.29791     & 2.028077    \\
					PR    & 2.116123    & 0.000651    & HER2  & 1.753906    & 2.369918    & ER    & 6.29791     & 2.028077    & ERBB2 & 3.023438    & 3.159443    \\
					PR    & 2.116123    & 0.000651    & HER2  & 1.753906    & 2.369918    & ERBB2 & 3.023438    & 3.159443    & ER    & 6.29791     & 2.028077   \\
					\hline
				\end{tabular}
			}
		\end{center}
		\label{table:gaussianfvar}
	\end{table}
	
	However, this problem can be overcome with a little compromise in the computation technique. In this study, we performed a stepwise computation technique to perform this analysis and obtain the threshold values. The gene expression values are standardized to treat them uniformly for threshold selection and difference in range of expressions can be nullified. For four biomarkers considered, we first set the threshold values at a particular level for the first three genes. The range of the 4th gene is divided into 100 equal parts, and each of them is considered as threshold values. The optimal threshold value for the 4th one is chosen from those 100 equidistant threshold values on the basis of the minimum p-value. To perform the analysis, we include all four binary variables generated from biomarkers in the CoxPH. Once the optimal threshold value is obtained for the 4th gene, it is kept fixed at that threshold value. Now, the same process is performed for the 3rd gene by keeping the starting threshold value for 1st and 2nd gene and the optimal threshold value for the 4th gene. The same goes for the 2nd gene and then for 1st gene consecutively. Thus, we obtain the optimal threshold value for each of the genes using starting threshold values of the previous genes and optimal threshold values of the following genes. As explained earlier that the biomarkers are dependent among themselves, so the threshold values obtained in this procedure will be dependent on the order of genes by which they are selected and their starting threshold values. To overcome this order dependency, we used all possible combinations of the genes and performed a similar analysis. The starting values are considered as the 1st, 2nd and 3rd quartile values of each gene. The output is shown in table \ref{table:genethreshold}. The optimal threshold values obtained using different starting values are mostly consistent.
	
	To verify the efficacy of the proposed procedure, it is vital to observe the effect of thresholding on survival predictions of patients. Shared frailty model has been utilised to observe the patient-specific randomness for the parts consisting of gene expression values below and above the thresholds. We implemented a proportional hazard model with shared frailty component, and comparison between the parts are made on the basis of frailty variances. The results of frailty variances considering threshold values obtained with 50th cutoff as starting value are shown in table \ref{table:gammafvar} \& \ref{table:gaussianfvar}. The differences between lower and upper frailty variance are evident. For ER and PR, the shared gaussian frailty variance for patients having gene expression value less than the threshold is higher than that of patients having higher gene expression value. For HER2, it shows the opposite characteristics. However, for ERBB2, the frailty variances are pretty close. For gamma shared frailty variance, the Only PR shows a similar output as gaussian frailty. For ERBB2, the frailty variance increases for higher expression level. However, both the frailty variances of the lower and upper part are very close to 0. For HER2, the frailty variance decreases for the higher part as opposed to the characteristics shown in gamma frailty. Foe ER, the frailty variances are pretty similar, and both of them are close to 0. 
	
	\section{Discussion}
	
	Biomarkers play a crucial role in the identification of disease progression. The treatment procedure of cancer is administered according to the response of corresponding biomarkers. In the case of breast cancer, several competing events may be present in the patient at the same time, and the occurrence of any one them may hinder the presence of others. The different biomarker statuses often portray contrasting effect on different events. However, it demands some statistical and medical justification for the splitting of continuous-valued gene expressions into different levels. 
	
	In this study, we have elaborated an efficient statistical procedure of dividing the gene expressions into categorical variables and validated the results by exploring the variability attached to the survival prediction at each of the biomarker levels. The survival predictions of different events in a particular biomarker level are likely to be correlated. This necessitates the use of correlated frailty where the correlation is translated through an additive frailty component consisting of biomarker level-specific frailty and event-specific frailty term. The variance of the frailty component is used as a criterion for thresholding of the gene expression. In our procedure, we used an additive frailty model under competing risk set-up and identified the event-specific frailty variance due to heterogeneity of biomarker levels. The event of interest is the occurrence of Distant Metastasis, and the competing event is the death from other causes. 
	
	In the proposed procedure, the gene expressions of 4 different biomarkers associated with breast cancer are studied. For each of the biomarker, the range of the expressions are divided into 100 equal parts and taking each one of them as the threshold levels, we have utilised a competing risk frailty model and obtained frailty variance associated with distant metastasis. The threshold value with maximum frailty variance provides the biomarker levels with most distinct effect on survival prediction. However, the threshold levels are obtained with the assumption that the biomarkers are independent, which is in contrary to the actual behaviour of the biomarkers. There is an association that exists among the biomarkers responsible for breast cancer, and the individual analysis of each of these biomarkers have lower power. Thus, a combined threshold level taking into considerations of all the biomarkers is necessary. Obtaining a consolidated frailty variance taking all the threshold levels for each biomarker is a difficult task. This problem can be overcome if we consider the corresponding p-values. The threshold levels with minimum p-values give maximum importance at each partition which will also provide minimum frailty variances. In our study, we have incorporated the frailty components as indicator variables in the model and obtained the threshold levels based on the minimum p-value criterion using p-value combiner methods. In the proposed procedure, we choose one indicator variable at a time keeping others fixed at certain threshold levels. Once an optimal threshold level is obtained, the corresponding biomarker level is set at that level, and the procedure is continued as before till we obtain threshold levels for all the biomarker levels. Also, to encounter the ordering problem, we work out the same procedure for all possible orders of the biomarkers and taking starting values as the quantiles. For validation, we applied the threshold levels for survival modelling and examined distinction in the survival prediction for each partition divided by the threshold levels. Also, this method effectively determines the optimal threshold levels in a sequential procedure instead of computing the minimum p-value for all possible combinations of thresholds. This dramatically reduces the computational time and minimises the multiple testing problem. Thus, using our proposed methodology, we obtain biomarker thresholding to clearly distinguish the survival effects for each partition and facilitate the treatment procedure. 
	
	The proposed procedure bears the potential for further improvements. In our study, we have created a criterion of choosing optimal threshold levels using p-value combiner methods to amalgamate the p-values corresponding to different correlated biomarkers. It will be beneficial to obtain a composite criterion associating the frailty variances for correlated biomarkers. Though we have utilised a competing risk model to obtain the frailty variances and incorporated the dependency structure among the events and biomarker levels, the prime interest is to examine the heterogeneity in survival prediction at different levels for only one event (i.e., Distant Metastasis in this study). It will be beneficial to extend the proposed procedure to examine heterogeneity in survival prediction for all the associated events. In this study, we partitioned the gene expressions on a dichotomous scale to identify the level separations. This can be extended on a polychotomous classification where the survival outcome differs at different levels of expressions. More precise results may be obtained using auxiliary information of the distributions of parameters which can be performed using bayesian techniques. 
	
	\section{Conclusion}
	
	The next generation of cancer treatment relies heavily on the proper identification of disease progression in terms of gene expressions. It is crucial to identify the distinct behaviour of survival outcomes and treatment responses at different biomarker statuses to provide the most effective treatment to the patients. We hope that the proposed procedure will motivate studies related to biomarker level classification and improve the patient treatment strategy. We are optimistic that the carefully designed prospective treatment regime will be assessed to take advantage of such clinical utility.
	
	\subsection*{Acknowledgments}
	Authors are also thankful to the Science and Engineering Research Board, Department of Science \& Technology, Government of India, for providing necessary support to carry out the present research work through project Grant No. EMR/2016/003305.

	\subsection*{Conflict of interest}
	
	The authors declare no potential conflict of interests.
	
	\subsection*{Supporting information}
	
	The detailed R\cite{R} programming code used in this article is available in the following link \url{https://github.com/souvikbanerjee91/modified-risk-detection}

\end{document}